\documentclass{article}
\usepackage{epsfig}
\usepackage{latexsym}

\begin{document}

\centerline{\sc \large Planet influence on the shape of the hosting star}
\centerline{\sc \large - ellipsoidal variations of tau Bootis}
\vspace{.5pc}
\centerline{W. Dimitrov}
\centerline{Astronomical Observatory of Adam Mickiewicz University}
\centerline{ul. S{\l}oneczna 36, 60-286 Pozna\'n, Poland}
\centerline{dimitrov@amu.edu.pl}

\vspace{2pc}

\abstract{This paper presents estimations on the possibility of detection of ellipsoidal variations
by means of measuring brightness of the star distorted by a close massive planet
using Wilson-Devinney method.
The problem was already discussed by Phafl et al. (2008) and earlier by Loeb \& Gaudi (2003).
The effect is well known in the case of
binary stars where it can produce light curves with amplitutudes of ellipsoidal
variations of about 0.1 mag for distorted stars.
For planets the effect is very small and usually less than 0.0001 mag.
The detection of an exoplanet, by searching for small amplitude ellipsoidal variations,
will be very difficult and affected by other photometric effects; however,
it maybe possible for some extreme cases.
Observations of ellipsoidal variations can provide additional constraints on the
model of the system.
Light curves for few star/planet systems have been calculated using PHOEBE eclipsing
binary software based on Wilson-Devinney method.
As an example of ellipsoidal variations the synthetic light curve for $\tau$Bootis is presented.
The amplitude of ellipsoidal variation is
0.01 mmag. The companion is massive (7.3 M$_{\mathrm{Jup}}$) and short-period hot Jupiter.}

\section{Introduction}\label{s1}

During the last decade we have observed a very fast development of extrasolar planet
research. The new methods of detection have been successfully used. The observational
techniques give us the possibility to detect more and more shallow effects like
transits of a planet in front of the hosting star, a direct imaging of planets and
measuring the chemical abundance in the planet atmosphere.
In this paper I will discuss the question whether we can observe light curve variations
of the hosting star distorted by the gravity field of a massive planet for some of the already known systems.
 The detailed analytical formulation of ellipsoidicity effect caused by
substellar companion is given by Pfahl et al. (2008).
A close massive companion of the star can produce tides analogue to the ones present
in close binary stars. The effect is of course much weaker and difficult for detection.
Actually, the presence of ellipsoidal variations among the transit candidates is treated as an evidence
of a non planetary behaviour of the eclipses.
Patzold et. al (2004) describe orbital period changes in planetary systems
caused by gravity tides induced by exoplanets.
Some short-period exoplanets, for example $\tau$Bootis, are probably in tidal lock
that demonstrates the significance of the tidal interaction.
We can model ellipsoidal variations using a code based on Wilson-Devinney (WD) method developed
for eclipsing binary stars. This method has already been used for modeling eclipses and
radial velocity curves of transiting planets - one of them being OGLE-TR-56 (Voccaro and Van Hamme 2005).

\section{Roche lobe and the massive planet}\label{s2}

I will use a standard Wilson-Devinney method (PHOEBE software, Pr{\v s}a and Zwitter 2005)
to describe the distortion
of the hosting star by its planetary companion. In the case of eclipsing binaries we have
a typical mass ratio of about $q=m_2/m_1 \sim 1$. For planetary systems, where $m_1>>m_2$
we usually have a mass ratio less than 0.01. The ellipsoidal light curve variations
in binary systems are dependent on the level of asymmetry of the distorted component.
We have higher amplitude of light variations for stars that significantly fill their
Roche lobes. For mass ratio of about 1 we observe distorted Roche lobes with high asymmetry
near Lagrangean point $L_1$. For $q<0.01$ the lobe of the star is almost spherical
(for slow rotation) with a small asymmetry near $L_1$.
The shape of the Roche lobe also depends on a rotational rate of the components. In
WD method we use the synchronicity parameter $f$,
defined as the ratio between the axial and orbital revolution period.
For high values of this parameter, $f\sim1$ (synchronous rotation),
we have a flattened star - this implies higher
asymmetry because of smaller distance between the stellar surface to $L_1$.
In table \ref{t1} we have listed values of amplitudes of ellipsoidal variations for two
synchronicity parameters $f$ of both the hosting star that fills its Roche lobe and for the MS star.
The amplitude also depends on the
gravity and limb darkening coefficients.
The main contribution in ellipsoidal variation amplitude is connected with the stellar shape
($>90\%$) and gravity brightening ($<10\%$).
We can expect notable ellipsoidal variations in two cases, considerable mass ratio or
high filling of the Roche lobe.
The distortion computed by the Roche model is slightly overestimated for evolved stars
because the mass is not concentrated in the barycenter but spread in a distorted star.

\begin{figure}
\begin{center}
\includegraphics[width=0.6\textwidth]{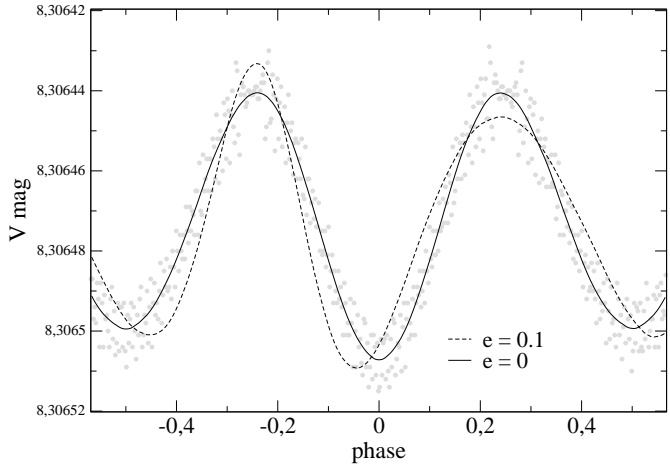}
\caption{Synthetic light curve of the hypothetical system for two eccentricities,
inclination of $79^{\circ}$ and synchronous rotation.
The gray dots present the
numerical noise for the circular orbit.}\label{f1}
\end{center}
\end{figure}

Let us take a 
system with a relatively distorted star characterized by
a mass ratio of 0.001,
a short orbital period of 1d and a semi-major axis of 5 R$_{\odot}$.
The main parameters of the system are close to those of OGLE-TR-56.
The mass of the planet is 0.0013 M$_{\odot}$ (1.3 M$_{\mathrm{Jup}}$).
The host star is slightly evolved with a mass of 1 M$_{\odot}$ and a radii of 1.3 R$_{\odot}$.
The obtained light curve has peak to peak amplitude of about 0.00006 mag.
The distorted star has the following dimensions:
$R_{\mathrm{side}}=1.32179$, $R_{\mathrm{point}}=1.32184$, $R_{\mathrm{back}}=1.32183$,
$R_{\mathrm{pole}}=1.30846$ in solar units.
Two curves are presented, one for circular and one for eccentric orbit (fig. \ref{f1}).
For a more evolved star with radii 2.5 R$_{\odot}$ we have peak to peak amplitude of 0.0005mag.

We have two minima of different depths, which is the
result of different surface temperature of the star near Lagrangean points
$L_1$ (lower) and $L_2$ (higher). Two maxima are different because of the
eccentric orbit that gives a variable separation of components for different phases.
This implies changes of radii/temperature of the star. We can also see another effect
of the orbital eccentricity - shifted minima and maxima.
The reflection effect was switched off, just as in the other modeled objects.
The numerical noise in LC is on the level of $10^{-5}$ for stellar surface grid raster
$n_{star}=60$ (maximal for PHOEBE). For the future aplications the exoplanet mode with denser
grid can be of use for modeling shallow effects.

\begin{figure}
\begin{center}
\includegraphics[width=0.6\textwidth]{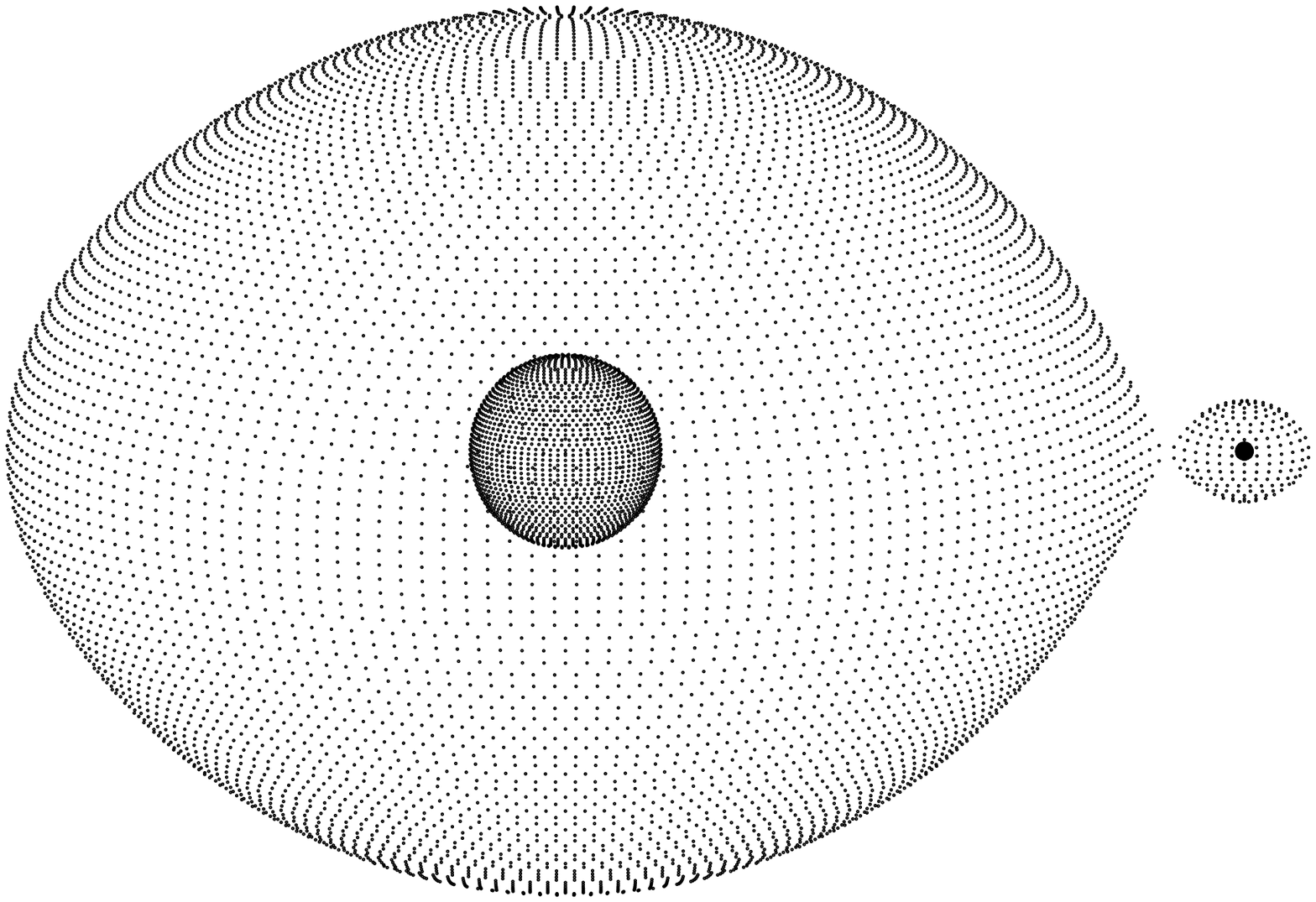}
\caption{Both components of the $\tau$Bootis system inside their Roche lobes for
inclination $80^{\circ}$, which presents better the flatness of the lobes.} \label{f2}
\end{center}
\end{figure}

\begin{table}[h]
\begin{center}
\caption{Maximal $V$ amplitude of ellipsoidal variations vs. mass ratio for the star
that fill its Roche lobe and for the main sequence star. Inclination is $90^{\circ}$,
component 2 is small and spherical - does not involve the curve shape. The host star
has one solar mass, the semi-major axis is 5 R$_{\odot}$ and the period $\sim 1.3d$.
}

\vspace{1pc}
\begin{tabular}{l|ll|ll}
\hline
mass ratio  & \multicolumn{2}{c}{ evolved } & \multicolumn{2}{c}{ main sequence }\\

   q        &   f=0.1 &\multicolumn{1}{c}{f=1} &   f=0.1 &  f=1   \\
\hline
   
   1       & 0.27    & 0.28  & 0.022   & 0.022   \\
   0.1     & 0.12    & 0.13  & 0.0025  & 0.0025  \\
   0.01    & 0.025   & 0.044 & 0.00025 & 0.00025 \\
   0.001   & 0.0045  & 0.013 &$10^{-5}$&$10^{-5}$\\
   0.0001  & 0.0004  & 0.0032& -       & -       \\
\hline
\end{tabular}
\label{t1}
\end{center}
\end{table}

\section{Ellipsoidal variations of known systems}\label{s3}

Among the discovered systems listed in \emph{exoplanet.eu}
data-base some stars have been checked for ellipsoidal variations
using WD simulations.
The star with higher amplitude is HD~41004B.
For the generation of its light curve the results of Zucker et al. (2004) were used.
Around the host star there is a massive companion (18.5 M$_{\mathrm{Jup}}$) orbiting,
the mass of which is above the limit of 13 M$_{\mathrm{Jup}}$ usually applied for
planets; therefore, we can classify this object as a brown dwarf.
The semi major axis of the orbit is very small 0.018 AU and the orbital period
is 1.3d.
Obtained LC presents variations with peak to peak amplitude of 0.0002 mag ($V$).
The theoretical amplitudes for other examined stars are: $\tau$Bootis - 0.00001,
OGLE-TR-56 - 0.00006 mag.

\begin{figure}
\begin{center}
\includegraphics[width=0.6\textwidth]{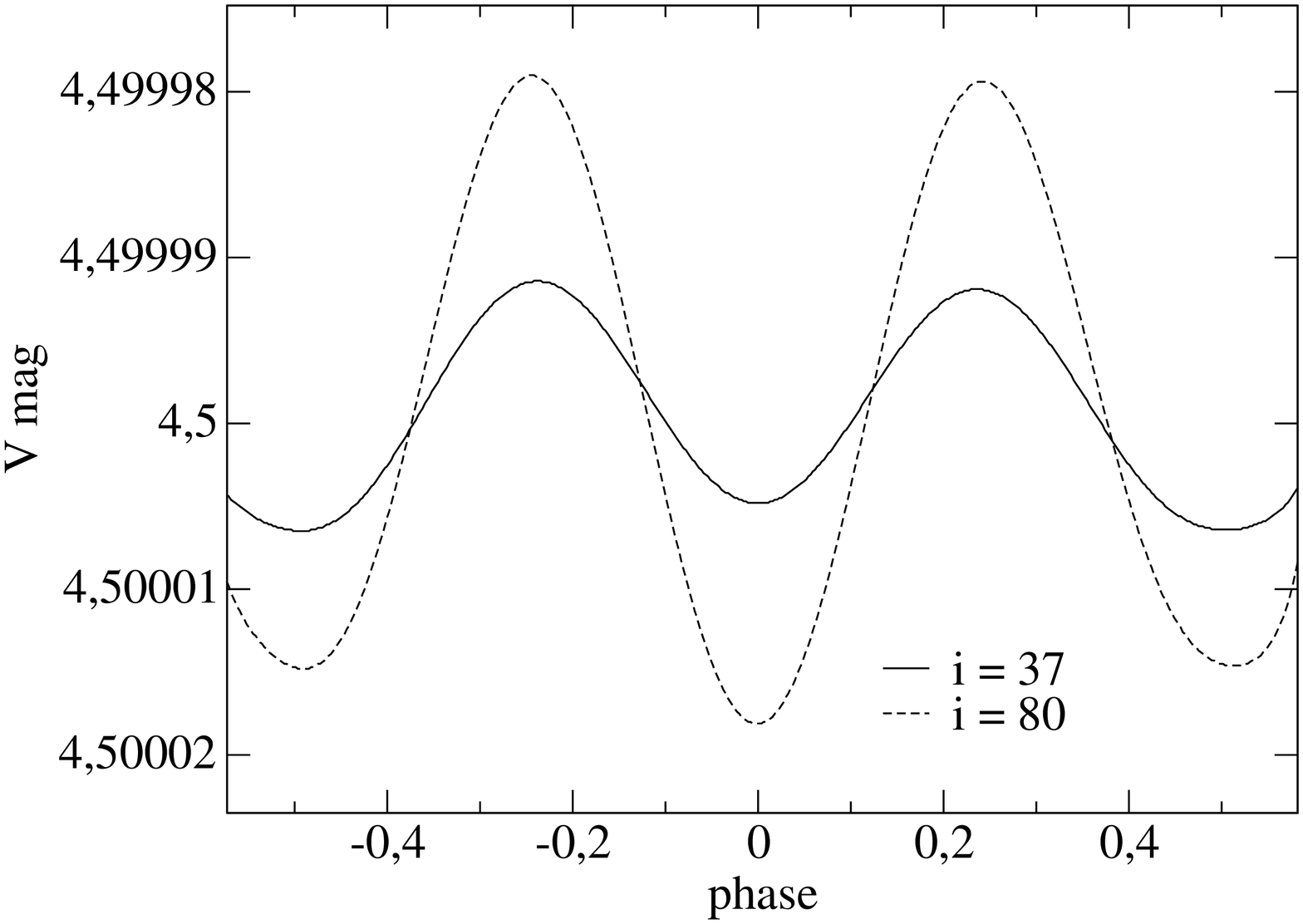}
\caption{Synthetic light curve of $\tau$Bootis A for two orbital inclinations.
The curves have been shifted for better comparison and smoothed to 
remove the "numerical noise".}\label{f3}
\end{center}
\end{figure}

The former - $\tau$Bootis - is a well studied system discovered in 1997 (Butler et al. 1997).
For the WD simulations system parameters (table \ref{t2}) listed in Leigh et al. (2003) were adopted.
This work presents spectroscopic search for reflected light. The projected orbital velocity
amplitude was found to be $K_p=97 \pm 8 km s^{-1}$ ($K_s=469 \pm 5 m s^{-1}$)
The most probable orbital inclination is $i=37 \pm 5^{\circ} $,
which implies planetary mass $M_p=7.28 \pm 0.83~$M$_{\mathrm{Jup}}$. The synthetic light curve
for two inclinations is presented in figure \ref{f3}.
The higher possible peak to peak amplitude of ellipsoidal variations without eclipses ($i=80^{\circ} $)
is $3 \cdot 10^{-5}$mag and for inclination $37^{\circ}$ we obtain $1 \cdot 10^{-5}$mag variability.
The eccentricity is small and unnoticeable on the light curve.
The shape of the star depends on the filling of the Roche lobe (fig. \ref{f2}).
During the evolution $\tau$Bootis will expand and fill equipotential surfaces with higher asymmetry,
which will increase the ellipsoidal variations (table \ref{t3}).

Future mass photometry space missions, for example Kepler (Christensen-Dalsgaard et al. 2006),
will be able to detect such level of variability.
The prospects of detection to be performed by Kepler mission have been discussed by
Pfahl et. al. (2008).
The differential photometry will reach precision of 20ppm for 12mag G2V star.
The detection of ellipsoidal variations will be affected by the other photometric
effects induced by planets - chromospheric activity, scattered light and, not connected
with the companion, for example spots and convection (the intrinsic stellar noise).
The MOST satellite has detected a variability on $\tau$ Bootis A probably induced by the planet
(Walker et al. 2008). The active region close to the sub-planetary point has been found. This
variable spot produces brightness variations with peak to peak amplitude of 1 mmag.
This effect is higher than ellipsoidal variations so their detection will probably be
impossible if the active region is stable and present in the future.
A similar effect was found on HD 179949 by Shkolnik et al. (2005). If such spots are typical
for 51 Peg systems, searching for ellipsoidal variations of hosting stars will be more
difficult.
The predicted level of scattered light for $\tau$Bootis by Green et al. (2003)
is two orders lower $1 \cdot 10^{-5}$mag and comparable with the ellipsoidal variations effect.

\begin{table}[h]
\begin{center}
\caption{Parameters of $\tau$Bootis system used for the synthetic curve
and obtained dimensions of the components for double tidal lock.}
\label{t2}
\vspace{1pc}
\begin{tabular}{ccc}
\hline
Parameter & star  & planet  \\
\hline

$i$        &      \multicolumn{2}{c}{      $37^{\circ}$         }\\
$q$        &      \multicolumn{2}{c}{      0.005                }\\
$A($R$_{\odot})$&  \multicolumn{2}{c}{      10.51 (0.0489 AU)    }\\
Period (d)&      \multicolumn{2}{c}{      3.31245               } \\
e &               \multicolumn{2}{c}{      0.018                }\\
$\omega$ &        \multicolumn{2}{c}{     $ 65^{\circ}$         }\\
\\
$\Omega$ & 7.12   & 1.93     \\
$x_{\mathrm{V}}$   & 0.54 & -    \\
\\
Mass$($M$_{\odot})$      & 1.42  & 0.007 (7.28 M$_{\mathrm{Jup}}$) \\
Radii$($R$_{\odot})$     & 1.48  & 0.12  (1.2 R$_{\mathrm{Jup}}$)\\
$T_{\mathrm{eff}}$ (K)      &         6360    & 1576 (estimated)  \\
\\
$R_{\mathrm{side}} ($R$_{\odot})$     & 1.480000 & 0.122414 \\
$R_{\mathrm{point}}($R$_{\odot})$     & 1.480034 & 0.122472 \\
$R_{\mathrm{back}} ($R$_{\odot})$     & 1.480028 & 0.122472 \\
$R_{\mathrm{pole}} ($R$_{\odot})$     & 1.477926 & 0.122393 \\

\hline
\end{tabular}
\end{center}
\end{table}


\begin{table}[h]
\begin{center}
\caption{Relation between the filling of the Roche lobe by the host star ($\Omega $) and
the ellipsoidal variations ($\Delta m$) for $\tau$ Bootis. The dimensions of the star (point \& side)
are given in units of semi-major axis.}
\vspace{1pc}
\begin{tabular}{lllll}
\hline
$\Omega $  &   radii R$_{\odot}$ & $\Delta m$ & $R_{\mathrm{point}}$ & $R_{\mathrm{side}}$ \\
\hline
 
 7       & 1.5    & $1 \cdot 10^{-5}$   & 0.143 & 0.143 \\
 3       & 3.6    & $2.5 \cdot 10^{-4}$ & 0.341 & 0.340 \\
 2       & 5.7    & $1.2 \cdot 10^{-3}$ & 0.542 & 0.541 \\
 1.6     & 8.0    & $0.005$             & 0.789 & 0.771 \\
 
\hline
\end{tabular}
\label{t3}
\end{center}
\end{table}


\section{Conclusions}\label{s4}
The future mass photometry missions can be used for detection
of exoplanet candidates by searching for effects weaker than transits like
induced spots, ellipsoidal variations or scattered light.
The WD code is already used for modeling transits of exoplanets. This
method can be usful for calculating weaker effects like
ellipsoidal variations or spots on the host star induced by the planetary companion,
and maybe applied for the scattered light after some modification of the code.
The method can also be useful for modeling the Rossiter-McLaughlin effect in
radial velocity curves.

\end{document}